\documentclass{emulateapj}
\usepackage{natbib}
\usepackage{epsfig}
\usepackage{graphicx}
\usepackage{subfigure}
\usepackage{float}
\usepackage{amsmath}
\usepackage{color}
\usepackage{amssymb}
\usepackage{amsthm}
\usepackage{mathrsfs}
\usepackage{amsfonts}
\usepackage{units}
\usepackage{mathtools} 
\usepackage[colorlinks,linkcolor=blue,anchorcolor=green,citecolor=blue]{hyperref}
\usepackage{upgreek}
\usepackage[normalem]{ulem}
\usepackage{verbatim}
\usepackage{tikz,xcolor,hyperref}

\definecolor{lime}{HTML}{A6CE39}
\DeclareRobustCommand{\orcidicon}{
	\begin{tikzpicture}
	\draw[lime, fill=lime] (0,0) 
	circle [radius=0.16] 
	node[white] {{\fontfamily{qag}\selectfont \tiny ID}};
	\draw[white, fill=white] (-0.0625,0.095) 
	circle [radius=0.007];
	\end{tikzpicture}
	\hspace{-2mm}
}
\foreach \x in {A, ..., Z}{%
	\expandafter\xdef\csname orcid\x\endcsname{\noexpand\href{https://orcid.org/\csname orcidauthor\x\endcsname}{\noexpand\orcidicon}}
}

\def\be{\begin{equation}}
\def\ee{\end{equation}}

\shorttitle{Magnetospheric origin of repeating FRB}
\shortauthors{Wang et al.}
\begin{document}

\title{On the magnetospheric origin of Repeating Fast Radio Bursts}
\author{Wei-Yang Wang\altaffilmark{1,2,3}\orcidA{}, Renxin Xu\altaffilmark{3,4,5}\orcidB{}, Xuelei Chen\altaffilmark{1,2,6}\orcidC{}}
\affil{$^1$Key Laboratory for Computational Astrophysics, National Astronomical Observatories, Chinese Academy of Sciences, 20A Datun Road, Beijing 100101, China}
\affil{$^2$University of Chinese Academy of Sciences, Beijing 100049, China}
\affil{$^3$School of Physics and State Key Laboratory of Nuclear Physics and Technology, Peking University, Beijing 100871, China}
\affil{$^4$Kavli Institute for Astronomy and Astrophysics, Peking University, Beijing 100871, China} 
\affil{$^5$Department of Astronomy, School of Physics, Peking University, Beijing 100871, China}
\affil{$^6$Center for High Energy Physics, Peking University, Beijing 100871, China}
\email{wywang@bao.ac.cn}
\email{xuelei@cosmology.bao.ac.cn}
\begin{abstract}
A bright radio burst was newly discovered in SGR 1935+2154, which exhibit some FRB-like temporal- and frequency-properties, suggesting a neutron star (NS)/magnetar magnetospheric origin of FRBs.
We propose an explanation of the temporal- and frequency-properties of sub-pulses of repeating FRBs based on the generic geometry within the framework of charged-bunching coherent curvature radiation in the magnetosphere.
The sub-pulses in a radio burst come from bunches of charged particles moving along different magnetic field lines. Their radiation beam sweep across the line of sight at different time, and those radiating at the more curved part tend to be seen earlier and at higher frequency. However, by 
considering bunches generated at slightly different times, we find there is also a small probability that the emission from the less curved part be seen earlier. 
We simulate the time--frequency structures by deriving various forms of the electric acceleration field in the magnetosphere.
Such structure of sub-pulses is a natural consequence of coherent curvature radiation from an NS/magnetar magnetosphere with suddenly and violently triggered sparks.
We apply this model to explain the time--frequency structure within specific dipolar configuration by invoking the transient pulsar-like sparking from the inner gap of a slowly rotating NS, and have also developed in more generic configurations.
\end{abstract}


\section{Introduction}
Fast radio bursts (FRBs) are mysterious millisecond-duration astronomical radio transients with large dispersion measures (DMs) and extremely high brightness temperatures \citep{Lorimer07,Keane12,Thornton13,Kulkarni14,Masui15,Petroff15,Spitler16,Petroff16,Chatterjee17}.
The excess of the Galactic DMs, and the localization of the host galaxies for several FRBs sources, shed light on the cosmological origins of FRBs \citep{Bassa17,Chatterjee17,Marcote17,Tendulkar17,Bannister19,Prochaska19,Macquart20,Marcote20}.
Some FRBs have been found repeating, and a very intriguing time--frequency structure was found \citep{CHIME19,Eight,Hessels19,Josephy19,Chawla20,Nine,Caleb20,Luo20} in some of the repeating FRBs.
For these bursts, each has several sub-pulses with narrow-band and different central frequencies, and arriving at the detector at different times.
For most of them, the time--frequency structure show a downward drifting pattern, i.e., the later-arrival sub-pulses have lower frequencies.
There are some possible upwards drifting tendencies shown in burst on MJD 58720, burst 191219A \& B, as well as in the bright and the faint sub-pulses in burst 191219 B of FRB 180916.J0158+65 \citep{Chawla20,CHIME20a}.
We have proposed a model by invoking a sudden-trigger-excited coherent curvature radiation in a neutron star (NS) magnetosphere to explain the downward drifting pattern \citep{Wang19}.

Very recently, a two-component bright radio burst with FRB-like temporal- and frequency-properties was detected by the Canadian Hydrogen Intensity
Mapping Experiment (CHIME; \citealt{Scholz20,CHIME20b}) and the Survey for Transient Astronomical Radio Emission 2 (STARE2; \citealt{Bochenek20a,Bochenek20b}) during the active state of the Galactic magnetar SGR 1935+2154, leading to the NS/magnetar-magnetospheric origin of FRBs. 
An X-ray burst detected by several X-ray instruments such as AGILE \citep{Tavani20}, Insight-HXMT \citep{Li20,Zhang20a,Zhang20b}, INTEGRAL \citep{Mereghetti20} and Konus-Wind \citep{Ridnaia20}, exhibits two hard peaks separated by $\sim30$ ms, in temporal coincidence with the radio burst event.
Deep searches by the Five-hundred-meter Aperture Spherical Telescope (FAST) revealed no FRB detection, during the epochs that 29 soft-$\gamma$-ray bursts were detected by Fermi \citep{Lin20}, suggesting SGR-associated radio burst event is very rare.
The FRB-like time--frequency structure of the radio burst is reminiscent to sub-pulse drifting which is a well-known phenomenon in some normal radio pulsars \citep{Rankin86} and explained as $\boldsymbol E\times \boldsymbol B$-induced drift in an NS magnetosphere \citep{RS75}.
The consecutive sparking process in the polar cap region of a normal radio pulsar would give rise to regular drifting of sub-pulses, while the repeating FRB sources and the SGR 1935+2154 bursts are more akin to a sudden and violent sparking process.
The models of that suddenly triggered mechanism in the magnetosphere of an NS/magnetar to account for FRBs, have been proposed by many authors (e.g., \citealt{Connor16,Cordes16,Dai16,Katz17,Zhang17,Wang18,Wadiasingh19}).

Motivated by the intriguing time--frequency structures in repeating FRBs and the FRB-like radio burst of SGR 1935+2154, in this paper, we expand the \cite{Wang19} model to explain the behaviors of suddenly and violently triggered sparks in the magnetosphere, by deriving the dynamics of particle bunches.
The model is demonstrated in \S\ref{sec:model}, and its applications in several scenarios are discussed in \S\ref{sec:application}.
The results are summarized in Section \S\ref{sec:sum} with some discussions.

\section{The magnetospheric model}\label{sec:model}

\begin{figure}
\includegraphics[width=0.48\textwidth]{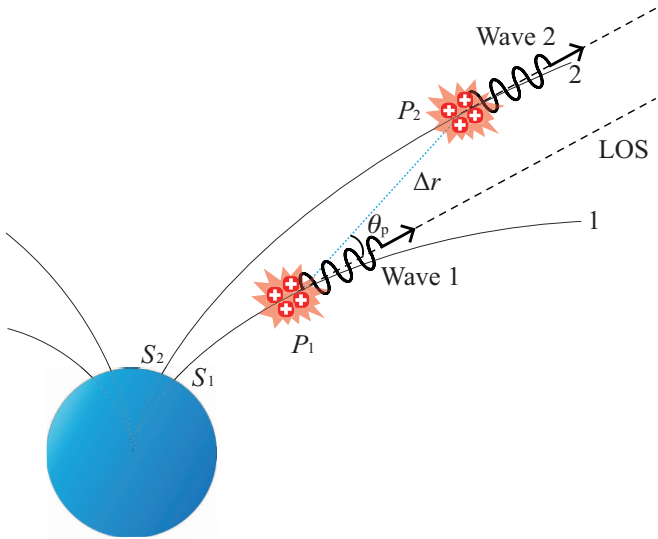}
\caption{\small{A schematic diagram of the NS/magnetar magnetosphere and radiation particle bunches. For instance, two magnetic field lines with charged bunches stream outflow are plotted. The dashed lines show the LOS. $P_1$ and $P_2$ denote the points where emissions can sweep the LOS at two magnetic field lines, i.e. two emitting points. Here, we assume that the curvature radius at $P_1$ is smaller than that at $P_2$. $S_1$ and $S_2$ denote the locations where the two bunches generation.
}}
\label{fig1}
\end{figure}

If the size of the charge particle bunch is less than half wavelength (for 1 GHz, $\lambda/2\sim10 $\, cm) or smaller, the phase of radiation from the particles are approximately the same, which allows the production of coherent radio emissions. In several models the FRBs are interpreted as coherent radio emission from charge particle bunches in pulsar magnetospheres \citep{Katz14,Kumar17,Lu18,Yang18,Lu20}.
We consider the coherent curvature radiation which frequency reads $\nu = (3/4\pi) \gamma^3 (c/\rho)$, where $\gamma$ is the Lorentz factor of the emitting particles, $c$ is speed of light and $\rho$ is the curvature radius.
The observed narrow band sub-pulses can be interpreted as a part of a multisegment broken power law spectra for bunching coherent curvature radiation from charge-separated clumps \citep{Yang20}.
The observed sub-pulses frequency is determined by both the geometric conditions and the dynamic of emitting particles.

\subsection{The geometry}

We evolve the previous model proposed by \cite{Wang19} which invokes bunching coherent curvature radiation in an NS magnetosphere.
A sudden trigger can let the NS becomes active, lasting several milliseconds or longer.
During the active state, many sparking events may occur, e.g., via sudden magnetic reconnection, or NS crust cracking.
Bunches of electron-positron pairs are created by these sparking mechanism, move along the curved field lines and process charge separation by an electric field parallel to the magnetic field lines.
The decouple and charge-separated bunches, e.g., positron clump, stream outwards along the open field lines, producing coherent curvature emissions.
We define the sub-pulses as emissions from these bunches which are produced by sparking events during the same trigger.
Magnetic field lines sweep across the line of sight (LOS) as the magnetosphere rotates.
Only the bunches which  arrive at their emission points with the LOS also happens to sweep there can be observed by us.
The observer can see the emission from several  sporadic bunches of neighboring magnetic field lines, as shown in Figure \ref{fig1}.

Basically, the motion of the  bunches along the field lines include both outflow and corotation with the NS in the magnetosphere simultaneously.
Therefore, one can calculate the time delay either via the projected horizontal motion, e.g., phase delay \citep{Wang19} or via the geometry of outflow (this work).
The observed time delay of the two sub-pulses is generically given by,
\be
\Delta t_{\rm obs}=\frac{d_2-d_1}{c}+t_{P_2}-t_{P_1},
\label{eq1}
\ee
where $d_1$ and $d_2$ are the distance of the two emitting points from us, and $t_{P_1}$ and $t_{P_2}$ are the times when the bunches arrived at their respective emitting points.
The observed emissions can be created by the bunches which are excited at different times.
{By deriving the motion of bunch outflow}, one can generally write
\be
t_{P_1}=t_{10}+\frac{s_1}{v_e}, \qquad
t_{P_2}=t_{20}+\frac{s_2}{v_e},
\label{eq2}
\ee
where $t_{10}$ and $t_{20}$ are the times when those bunches were generated, $s_{1}$ and $s_{2}$ are the distances that the bunches travelled from the generation points to the emitting points along the field lines, and $v_e=\beta_e c$ is the velocity of the bunch particles.
Combined with Eq. (\ref{eq1}) and Eq. (\ref{eq2}), one can obtain 
\be
\Delta t_{\rm obs}=\Delta t_{\rm geo}+t_{20}-t_{10},
\label{eq3}
\ee
where $\Delta t_{\rm geo}$ is the geometric time delay, which can be written as
\be
\Delta t_{\rm geo}=\frac{s_2-s_1}{v_e}-\frac{\Delta r\cos\theta_p}{c},
\label{eq4}
\ee
where $\Delta r\cos\theta_p$ is the projection of the distance between the two emitting points in the direction of the LOS.

According to Eq. (\ref{eq3}), if two bunches are generated simultaneously, their emissions would be observed at different epochs.
In this case, the emission from the more curved part (line 1) is seen earlier than that from the less curved (line 2).
On the other hand, the emission from line 2 may arrive the detector at an earlier time if the bunch was generated much earlier.
For the geometry-dependent plasma emission mechanism, one would observe different time--frequency structures from some sub-pulses.

\subsection{Acceleration of charged particles}

An electric field $E_{\parallel}$ parallel to the B-field that sustain the acceleration of charged particles is required, otherwise the particles will cool down very rapidly \citep{Kumar17}.
The $E_{\parallel}$ can accelerate bunches of electrons and positrons to ultra-relativistic velocities moving in opposite directions.
The acceleration of the charged bunches can be written as
\be
N_eE_{\parallel}eds-Ldt=N_em_ec^2d\gamma,
\label{eq5}
\ee
where $N_e$ is the number of positrons in the bunches, $L$ is the total luminosity of the radiation, $m_e$ is the positron mass.
In order to sustain a constant Lorentz factor within a labframe duration of $\gamma^2/\nu$, i.e., balance between acceleration and radiation cooling, one requires an $E_{\parallel,0}$ \citep{Wang19}
\be
E_{\parallel,0}\simeq\frac{\gamma m_{\rm e}c}{(et_{\rm cool})}\sim3.1\times10^7\nu_9^2N_{\rm e, 23} \gamma_{2}^{-2}\,\rm esu,
\label{eq:Epara}
\ee
where we adopt the convention $Q_x=Q/10^x$ in cgs units.
If the strength of $E_{\parallel}$ is much stronger than $E_{\parallel,0}$, the Eq. (\ref{eq5}) can be approximated as
\be
E_{\parallel}e\beta_e dt\simeq m_ecd\gamma.
\label{eq6}
\ee

Many trigger mechanisms have been proposed to explain FRBs (see \citealt{FRBmodel} for review).
One possible scenario is that the $E_{\parallel}$ may be triggered together with the bunch generation.
We investigate a list of possible trigger mechanisms that might create $E_{\parallel}$ in the magnetosphere.
The mechanisms include:\\
(1) Crustal-deformations-triggered field via Alv{\'e}n wave in the charge starvation region \citep{Kumar20}.
The Alv{\'e}n waves can be created by sudden crustal motion, starquake or emergence of magnetic flux tubes (e.g., \citealt{Wang18,Wadiasingh19}), and then propagate in the inhomogeneous magnetosphere.
The timescale for this process can be estimated as $\sim R/v_{\rm A}\approx0.3\,\rm s$, where $R$ is the radius of the NS surface and $v_{\rm A}$ is the Alv{\'e}n speed adopted as $0.01c$.
An electric field is formed in the charge starvation region, since the plasma is insufficient to supply the current needed by the Alv{\'e}n wave.
The acceleration field is described as \citep{Kumar20}
\be
E_{\parallel}=E_{\rm d}\sin[\omega_{\rm aw}(s/c-t)],
\label{alven}
\ee
where $\omega_{\rm aw}$ is the angular frequency of the Alv{\'e}n wave and $E_{\rm d}$ is the field that provides displacement current given by
\be
E_{\rm d}=-(k_{\rm aw\perp}/k_{\rm aw\parallel})\delta B\propto r^{-3}.
\ee
If the Alv{\'e}n wave oscillates not so rapidly, e.g., the term of phase $\sim-10^{-2}\omega_{\rm aw,5}t_{-3}\gamma_2^{-2}$, the acceleration for charged bunches is mainly determined by the power law component; \\
(2) A self-induced field in the twist-current-carrying bundle \citep{Beloborodov07}.
When the star undergoes a sudden crust quake, the motion of footpoints can make the outer magnetosphere twist up.
During this process, the ejected currents flow along the field lines to the exterior of the star and come back at other footpoints, supporting the twisted magnetic field.
An $E_{\parallel}$ is generated by the self-induction of the current, which can be written as
\be
E_{\parallel}\simeq\frac{c\tau B\sin^2\theta\Delta\phi}{r}\propto r^{-4},
\label{twistE}
\ee
where $\tau$ is the timescale of the current dissipation, $B$ is the magnetic field strength and $\Delta\phi$ is the twist angle;\\
(3) The ``cosmic comb'' model \citep{Zhang17}.
A plasma stream from a nearby source (e.g., binary companion or massive black hole, etc.) interacts the NS with the ram pressure overcoming the magnetic pressure.
The sudden distortion of the magnetosphere can derive the number density deviating the original Goldreich--Julian value, forming bunches of charged particles in a locus of field lines.
An $E_{\parallel}$ is created due to the sudden compression of the Goldreich--Julian density.
For an easy description, we relate deviation of the net charge density with the original Goldreich--Julian density, i.e., $\xi B/(Pce)$, where $\xi$ is the compression factor and $P$ is the period.
Thus, one has $\nabla\cdot E_{\parallel}\simeq4\pi \xi B/(Pc)$.
Combined with the boundary condition of $E_{\parallel}=0$ at $r\gg cP/(2\pi)$, the acceleration field is given by
\be
E_{\parallel}\simeq\frac{2\pi \xi Br}{Pc} \propto r^{-2}.
\label{comb0}
\ee
On the other hand, in order to derive a constant Lorentz factor, the required electric field reads \citep{Wang19}
\be
E_{\parallel,0}\simeq\frac{8\pi \mu\eta_cBr}{27\gamma^2\lambda} \propto r^{-2},
\label{comb}
\ee
where $\mu$ is the normalized fluctuation of electrons and $\eta_c$ is a parameter describing the cross section.

The $E_{\parallel}$ proposed in these models can all be written in a power-law form, but differ in their indices.
Another possible scenario for the $E_{\parallel}$ is that it is the acceleration field in the slot gap above the pulsar polar caps \citep{Arons83}.

\subsection{Drifting parttern}
Similar to normal pulsars, the drifting pattern of FRBs can reflect the emitting conditions changing with the locations at magnetosphere (e.g., ``radius-to-frequency mapping'', \citealt{Lyutikov20}).
The drift rate can be written as
\be
\dot{\nu}=\nu\left[3\frac{\Delta\gamma}{\gamma\Delta t_{\rm obs}}-\frac{\Delta \rho}{\rho\Delta t_{\rm obs}}\right].
\label{eq7}
\ee
If $E_{\parallel}$ is very close to $E_{\parallel,0}$, the Lorentz factors of the bunches would be the same for each and do not evolve significantly as they stream out along the field lines, therefore the drift rate can be simplified as
\be
\dot{\nu}=-\nu\frac{\Delta \rho}{\rho\Delta t_{\rm obs}}.
\label{eq8}
\ee
The drift rate is mainly determined by the change of curvature radius of the emitting points.
Generally, the emitting points swept the LOS earlier emit curvature radiation in the more-curved part of the field lines, resulting in downward drifting patterns.
On the other hand, if the two bunches are not triggered simultaneously, e.g., the bunch at line 2 is generated earlier than that at line 1, from Eq. (\ref{eq3}), the wave from the line 2 could be observed earlier when $|t_{20}-t_{10}|$ is longer than $t_{\rm geo}$, so that upwards drifting pattern will be seen.
For the scenario of $E_{\parallel}\gg E_{\parallel,0}$, the drift rate can be influenced by the difference of $\gamma$ for the bunches at different field lines.
A complex drifting pattern may be caused by the complicated $E_{\parallel}$.

\section{Applications}\label{sec:application}

We apply this radiation model to two scenarios.
In the first scenario, the magnetic field configuration is a simple dipole, pulsar-like sparking processes generate bunches of charged particles at the inner gap region, FRBs are subsequently produced in the polar gap region at the open field lines.
The second scenario is more generic, here we do not specify the trigger mechanism, just give a generic field line configuration.
The sparking process can be caused by either internal or external excitation.

\subsection{Polar gap sparking in a dipole field}

\begin{figure}
\includegraphics[width=0.48\textwidth]{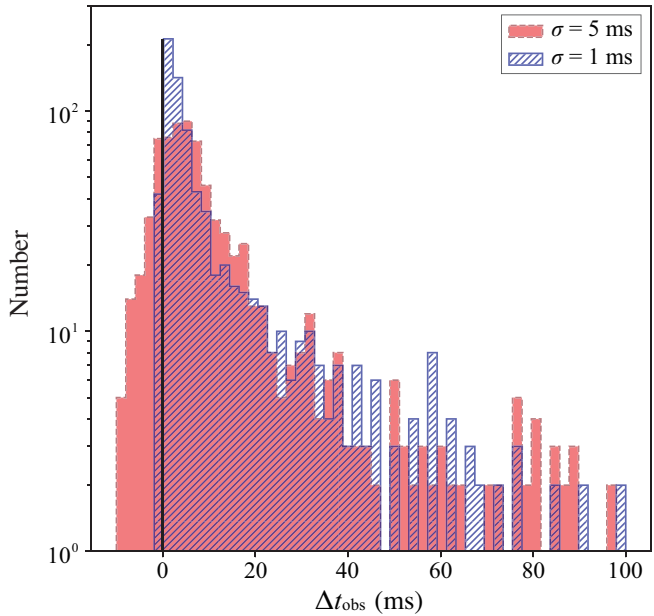}
\caption{\small{Histograms of estimated $\Delta t_{\rm obs}$ from the Eq. (\ref{eq3}). The time delay due to the un-simultaneous sparking is assumed as normally distribution with $\mu = 0$ and $\sigma=1\,\rm ms,$ (blue) $5\,\rm ms$ (red). The $\Delta t_{\rm geo}$ is simulated by making the $\sin\theta_s$ uniformly distributed when $\theta_s$ varies in $0-0.02$. The vertical black solid line divides events with positive/negative values. 
}}
\label{fig2}
\end{figure}

\begin{figure*}
\begin{centering}
\includegraphics[width=0.98\textwidth]{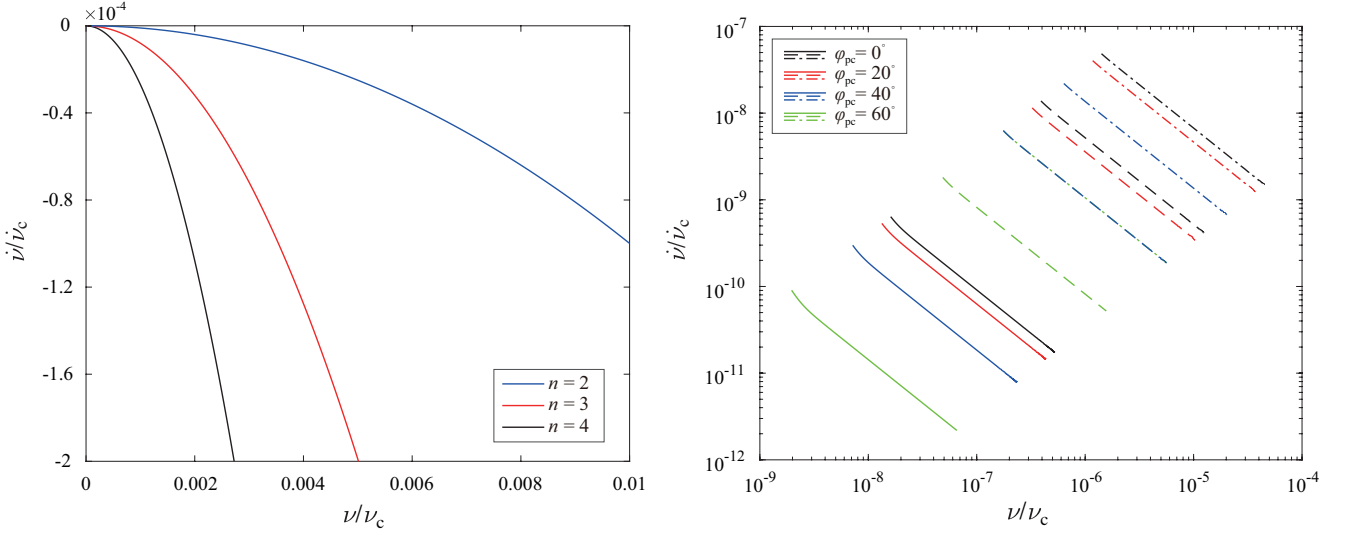}
\caption{\small{ The drift rate as a function of the emission frequency. Both the frequency and drift rate are normalized with their characterized parameters. Left: Simulated drifting patterns for power-law-distributed $E_{\parallel}$ are plotted in solid lines with $n=2$ (blue), $n=3$ (red) and $n=4$ (black); Right: Same as the left but for the $E_{\parallel}$ in the slot gap with $\theta_c=0.02$. The patterns with different inclination angles are plotted in solid lines ($\alpha=10^\circ$), dashed lines ($\alpha=30^\circ$) and dashed-dotted lines ($\alpha=50^\circ$) for $\phi_{\rm pc}=0^\circ$ (black), $\phi_{\rm pc}=20^\circ$ (red), $\phi_{\rm pc}=40^\circ$ (blue) and $\phi_{\rm pc}=60^\circ$ (green). 
}}
\label{fig3}
\end{centering}
\end{figure*}

We consider a scenario similar to the polar gap sparking of the normal radio pulsars \citep{RS75}. 
The mechanism we envisage here is a sudden and violent sparking process to produce bunches at the stellar surface.
When the sudden sparking perturbs the magnetic field to deviate from the regular state significantly, bunches of charged particle will produce coherent curvature radiation in a locus of field lines \citep{Yang18}.
Multiple bunches emitting in adjacent field lines traveling with a similar initial Lorentz factor as the perturbation propagates outwards as Alfv{\'e}n waves.

In the spherical coordinates $(r,\,\theta,\,\phi)$ with respect to the magnetic axis, a magnetic field line of the dipolar configuration can be described as
\be
u=\frac{R\sin^2\theta}{r}=\sin^2\theta_s,
\label{eq:dipole}
\ee
where $u$ is a dimensionless constant and $\theta_s$ is the angle of the foot point for each field line at the stellar surface. 
For a field line characterized by $u$, the distance along the field from $\theta_0$ to $\theta$ is given by
\be
s=\frac{R}{u}\int^{\theta}_{\theta_0}\sqrt{1+3\cos^2\theta}\sin\theta d\theta.
\label{eq9}
\ee
The polar gap region is enclosed within the last open field lines, which has a polar angle $\theta_c=0.1(P/10\,{\rm ms})^{-1/2}$.
In order to have coherent emissions, the bunch opening angle should be smaller than $1/\gamma$, suggesting $P > 0.3\,{\rm s} (\gamma/300) \Delta t_{\rm obs}$ \citep{Wang19}.
Line 1 and line 2 are at the same plane approximately, i.e., the rotating effects \citep{Lyutikov20} can be neglected.
Therefore, we obtain $\theta_c < 0.02(\gamma/300)^{-1/2}$.
Within the region $\theta\lesssim0.5$, Eq. (\ref{eq9}) can be approximated as
\be
s\simeq\frac{2R}{u}(\cos\theta_0-\cos\theta)\simeq r.
\label{eq10}
\ee
From Eq. (\ref{eq4}), the geometric time delay can be written as
\be
\Delta t_{\rm geo}\simeq\frac{\Delta r}{c}(1-\cos\theta_p+\frac{1}{2\gamma^2}).
\label{eq11}
\ee
The curvature radius of the field line can be also described as $\rho
\approx 4r/(3\sin\theta)$.

Assuming that the Lorentz factor is a constant, and first consider the bunches are generated at the same time. Combined with Eq. (\ref{eq8}) and (\ref{eq11}), the drift rate can be calculated as
\be
\dot\nu=-\nu^2\frac{16\pi}{9\gamma^3(1-\cos\theta_p+1/2\gamma^{-2})\sin\theta}.
\label{eq12}
\ee
The luminosity of the coherent curvature radiation reads $N_e^2\delta p_e$, where $\delta p_e$ is the luminosity for one electron/positron.
From Eq. (\ref{eq5}), to sustain a constant Lorentz factor, one requires the acceleration field obeys a function of $E_{\parallel,0}\propto\rho^{-2}$.

We then consider that sparks are produced at different times.
This may be caused by a rough surface of pulsar \citep{Lu19}, and the binding energy of particles on stellar surface need to be very high (e.g., in a bare neutron star with no gaseous atmosphere sits on top of the crust \citep{Turolla04} or solid strangeon star \citealt{Xu99,Yu11}).
The distance of the two emitting points can be written as
\be
\Delta r=\frac{-2R\sin^2\theta\cos\theta_s\Delta\theta_s}{\sin^3\theta_s}.
\label{eq13}
\ee
Thus, the geometric time delay is estimated as
\be
\Delta t_{\rm geo}\simeq6.7\times10^{-2}\,{\rm ms}~\theta^2\Delta\theta_s\sin^{-3}\theta_s(1-\cos\theta_p).
\label{eq14}
\ee

To assess the overall time--frequency structure of the bursts in this model, we make a simulation of the bursts. 
We adopt a uniformly distributed $\sin\theta_s$ in the range of $0-0.02$, $\theta=0.1$ and $\Delta \theta_s=0.01$ rad in the following simulation.
The term of $1-\cos\theta_p\simeq1$ is assumed.
The term of $t_{20}-t_{10}$ follows the normal distribution with $\mu = 0$ and $\sigma=1\,\rm ms,\,5\,\rm ms$, respectively.
Based on Eq. (\ref{eq3}), we plot the simulated distribution of the observation interval time $\Delta t_{\rm obs}$ with 1000 samples, as shown in Figure \ref{fig2}.
The temporal properties of the simulated patterns are consistent with the observations from most repeating FRBs.
The vertical black solid line divides positive/negative slope values.
According to our simulation, in most cases one would observe a downward drifting pattern, though there are also cases in which an upward drifting will be seen.
As the value of $\sigma$ increases, the probability of upwards drifting will also increase and gradually tends to 50\%.

Different trigger mechanisms may lead to various forms of accelerating electric field $E_{\parallel}$.
For the mechanisms mentioned in Sec. 2.2, it has a power law form, i.e., $E_{\parallel}=E_0\left(r/R\right)^{-n}$, when $E_{\parallel}\gg E_{\parallel,0}$.
Different power law indices reflect different trigger mechanisms.
From Eq. (\ref{eq7}), the drift rate can be written as
\be
\dot\nu=\frac{9E_0^3e^3R\sin\theta}{16\pi m_e^3c^4(1-\cos\theta_p+1/2\gamma^{-2})}C_n(\eta),
\label{eq15}
\ee
where $\eta=r/R$ which is assumed to be much larger than 1 and $C_n(\eta)$ is given by
\be
C_n(\eta)=\begin{dcases}
\frac{[(2-3n)\eta^{1-n}+1](\eta^{1-n}-1)^2}{(1-n)^3\eta^2}, &n\not=1\\
\frac{3[{\rm Log_e}(\eta)]^2-[{\rm Log_e}(\eta)]^3}{\eta^2}, &n=1
\end{dcases}.
\label{eq:K(x)}
\ee
The drift rate and the emission frequency are normalized in $$\dot{\nu_c}=(9E_0^3e^3R\sin\theta)/[16\pi m_e^3c^4(1-\cos\theta_p+1/2\gamma^{-2})],$$ 
and 
$$\nu_c=9E_0^3e^3R^2\sin\theta/(16\pi m_e^3c^5).$$
In Figure \ref{fig3}, we show the simulated drift rate as a function of the emission frequency with different value of $n$.
More generally, the sign of the drift rate depends on $n$.
As shown in Eq. (\ref{eq:K(x)}), for the acceleration electric fields decrease with $r$ rapidly ($n\gtrsim1$), the growth of the Lorentz factor at higher height is slow, so that frequencies may drift from high value to low.
If $n$ is much larger than 1, the drift pattern is approximately ${\dot\nu}\propto-\nu^2$.
For $n\lesssim1$, the Lorentz factor at higher height is much larger, therefore both downwards and upwards drifting are possible.

For the slot gap, $E_{\parallel}$ has a more complex form.
The slot gap region is bounded by the last open field lines and  the magnetic co-latitude lines.
In this region, the combination of geometrical screening and the effect of frame dragging can give rise to a regime of extended acceleration for charged particles, with the
acceleration rate 
\be
\frac{d\gamma}{ds}\simeq A\left(\frac{\kappa}{\eta^4}\cos\alpha+\frac{\theta_c}{4}\eta^{-1/2}\sin\alpha\cos\phi_{\rm pc}\right),
\label{eq16}
\ee
where $A$ is a factor, $\kappa=0.15$ is the general relativistic parameter entering the frame-dragging effect, $\alpha$ is the inclination angle and $\phi_{\rm pc}$ is the magnetic azimuthal angle \citep{Muslimov03,Muslimov04}.
Here, we set $E_0=Am_ec^2/e$.
From Eq. (\ref{eq7}), the drift pattern reads
\be
\nu=\nu_c\left(-\frac{\kappa}{3}\eta^{-10/3}\cos\alpha+\frac{\theta_c}{12}\eta^{-5/6}\sin\alpha\cos\phi_{\rm pc}\right)^3,
\label{eq17}
\ee
and
\be
\dot{\nu}=\frac{\dot \nu_c}{\nu_c}\left(\frac{d\nu}{d\eta}\right).
\label{eq18}
\ee
In Figure \ref{fig3}, we plot the drifting patterns for bunches accelerated by the $E_{\parallel}$ in the slot gap with different parameters.
All the patterns show upwards drifting structures because the $E_{\parallel}$ decrease not so rapidly that Lorentz factors at high height are large, leading to high-frequency waves.

Both mechanisms give rise field strength  decay with heights, so that the $E_{\parallel}$ at higher region is closed to $E_{\parallel,0}$.
Thus, the drifting pattern tends to ${\dot\nu}\propto-\nu^2$ at lower-frequency bands.

\subsection{A generic field configuration}

In this generic model, we consider that the magnetic fields deviate from a simple dipole.
For instance, a possible scenario is a highly twisted magnetosphere of a magnetar, e.g., \cite{Thompson95}.
The pulsar-like sparking process invoked here is similar with the case discussed in Section 3.1 but electron-positron pairs are created via two-photon-pair production.
Another scenario is that the magnetosphere is combed by a plasma stream \citep{Zhang17}.
The sudden trigger can make the number density deviation, forming sparks in a locus of field lines and providing acceleration field of the charged bunches.

For the generic model, the magnetic field can be generally written as ${\boldsymbol B}(B_{r},\,B_\theta,\,B_\phi)$. 
Then we implement the following procedure:\\
1) Give the geometry of field lines via $B_{r}/B_\theta=dr/(rd\theta)$ and $B_\theta/B_\phi=d\theta/(\sin\theta d\phi).$\\
2) Find the positions of bunch generation and emitting point at each field lines.\\
3) Calculate the distances that the bunches travelled from the generation points to the emitting points along the field lines.\\
4) Derive the observed time delay from Eq. (\ref{eq3}).\\
5) Find $E_{\parallel}$ in the magnetosphere during the trigger.\\
6) Calculate the Lorentz factors from Eq. (\ref{eq5}).\\
7) Combined with the geometry of the field configuration and the dynamic of emitting bunches, one can derive the drifting pattern from Eq. (\ref{eq7}).\\

\section{Summary and discussion}\label{sec:sum}

We propose a generic geometrical mechanism to explain the temporal- and frequency-properties of sub-pulses of repeating FRBs within the framework of coherent curvature radiation in the magnetosphere of an NS/magnetar.
The bunches of charged particles are produced by suddenly and violently triggered sparks from the magnetosphere, and stream outwards along the open field lines.
The observed sub-pulse emission can then interpreted as the emission from several sporadic bunches of neighboring magnetic field lines.
This model is applied to explain the time--frequency structure within a specific dipolar configuration by invoking sparking from the inner gap of a slowly rotating NS, and also developed in more generic configurations.
For an NS with sudden and violent triggered sparking, such structure of sub-pulses is a natural consequence of coherent curvature radiation from sporadically seen bunches.
We further argue that such structure could be regarded as an evidence of magnetospheric origin of FRBs.
We are expecting to test this FRB model of magnetospheric origin in the future with advanced facilities, especially by the FAST \citep{FAST}, which has great advantages in detecting faint radio bursts, and the CHIME in detecting more FRB samples \citep{CHIME18}.

The temporal- and frequency-properties of sub-pulses are supposed to be strongly related to the geometry of the magnetosphere and the dynamics of charged particles within it.
In principle, a complicated time--frequency structure could originate from a very complex magnetosphere or the acceleration electric field.
The simplest scenario is that these bunches are triggered simultaneously, and move with a nearly constant Lorentz factor, suggesting the potential observation of a $\dot{\nu}\propto-\nu^2$ relation.
The interval time between two adjacent sub-pulses should be longer for lower-frequency emissions because they are emitted at higher height.
Sub-pulses with short interval times, are more likely to be seen in higher frequency bands.

A more complex case is that the bunches are generated with slightly different times.
We simulated the observed intervals of sub-pulses in a dipole magnetic configuration.
We found in most cases, the emissions from more curved parts of the field lines are seen earlier than those from the less curved part.
As shown in Figure \ref{fig2}, 
more events with $\Delta t_{\rm obs}<0$ can occur for larger $\sigma$, suggesting that the long-duration trigger mechanism is more likely to generate upward drifting events. 
For a long enough trigger event, the chance of observing upwards and downwards drifting should be the same.
Our model model predicts that most FRBs would have downwards drifting sub-pulses, but there are also upward drifting events, and one would more likely to observe these from FRBs with long-duration.

Most repeating FRBs have same order of drift rates at $400-800$ MHz \citep{CHIME19,Eight,Nine}, suggesting that FRB sources most likely have similar magnetospheres.
This similarity is also enhanced by the measurement of drift rates at $\sim1$ GHz for FRB 121102 and FRB 181123 \citep{Zhu20}.
For FRB 121102, the drift rates of bursts detected at higher frequency band, are generally larger than those at lower frequency band \citep{Hessels19,Josephy19,Caleb20}, matching the scenario of a rapidly decreasing $E_{\parallel}$.
 
A pulsar or magnetar, when excited by a sudden trigger creating sporadic sparks, may be seen to have such sub-pulse structures within some single transient pulses.
The sudden trigger for a normal pulsar may be related to an abrupt crust cracking, which should be accompanied with glitches \citep{Ruderman98}.
Magnetars can create sparks via magnetospheric twist and becomes active \citep{Beloborodov13,Wadiasingh20}. It is associated with some X-ray bursts, which is consistent with observations of SGR 1935+2154 \citep{Bochenek20a,Bochenek20b,Li20,Mereghetti20,Ridnaia20,Scholz20,CHIME20b,Zhang20a,Zhang20b}.
However, the luminosity of the radio burst event of SGR 1935+2154 is $1-2$ orders of magnitude smaller than that inferred of typical FRBs.
FRB may have similar trigger mechanism but differ in energy budget.
In general, the $E_{\parallel}$ proposed in most FRB models are related to the trigger mechanism, which 
could be revealed by the observation of the time--frequency structure.
We also consider the $E_{\parallel}$ in a pulsar slot gap, however, only upwards drifting pattern is expected in this case.

The radio bursts of SGR 1935+2154 are separated by $\sim30$ ms \citep{Scholz20,CHIME20b}, which is larger than the bursts' duration.
We note that there is observational differences between conventional FRBs and the radio bursts of SGR 1935+2154, but this could be the result of different activity levels in our sparking scenario, to be high for the former (so that charged-bunching coherent radiation beams would overlap) but low for the latter (separated by $\sim30$ ms). 
Bunches flow along the neighboring field lines and emit at different heights. 
Therefore, one may see temporally separated sub-pulses due to the low active level and the geometric effect.
The $\sim30$\,ms-separated two peaks detected from the X-ray burst lasting 0.4 s \citep{Li20,Mereghetti20,Ridnaia20,Tavani20,Zhang20a,Zhang20b}, are consistent with the separation between the CHIME events, indicative of that the separated radio bursts are two sub-pulses during the 0.4 s-lasting trigger.
From another point of view, for a pulsar with spin period of $\sim 3$ s, the angle between the magnetic axis and the field direction of the last-open field line could be $\sim 2.3^\circ$ (at an attitude of 100 km), and time delay of two sub-pulses detected could be order of $\leq 40$ ms.
The sub-pulse emissions should be at least within one pulsar's period.
This would be tested by future observation of periodicity, if the sudden trigger can last at least one period.

For a slow rotating pulsar with ideal dipolar configuration, the polarization angle is generally flat but evolve at near $\phi=0$ (see the RVM model, \citealt{RVM}). 
More complicated magnetic field configurations may produce more complex polarization angle evolution curves. 

\acknowledgments
{{We thank Xueying Zheng for helping to simulate the possibility of upwards and downwards drifting. We are grateful to Dr. Wenbin Lu, Dr. Rui Luo, Dr. Yuan-Pei Yang, Prof. Bing Zhang and Dr. Enping Zhou and an anonymous referee for helpful comments and discussions. W.-Y. W. and X. L. C. acknowledge the support of the NSFC Grants 11633004, 11653003, the CAS grants QYZDJ-SSW-SLH017, and CAS XDB 23040100, and MoST Grant 2018YFE0120800, 2016YFE0100300, R. X. X. acknowledges the support of National Key R\&D Program of China (No. 2017YFA0402602), NSFC 11673002 and U1531243, and the Strategic Priority Research Program of CAS (No. XDB23010200).}}

\end{document}